\documentclass[12pt,preprint]{aastex}

\begin{document}


\title{A Simple and Accurate Model for Intra-Cluster Gas  }


\author{Jeremiah P. Ostriker\altaffilmark{1} and Paul Bode} 
\affil{Department of Astrophysical Sciences, Peyton Hall, Princeton
University, Princeton, NJ 08544}
\and
\author{Arif Babul\altaffilmark{2}}
\affil{Department of Physics \& Astronomy, University of Victoria,
Victoria, BC, V8P 1A1, Canada }


\altaffiltext{1}{Institute of Astronomy, Madingley Road, Cambridge CB3
0HA, UK}
\altaffiltext{2}{Astrophysics, University of Oxford, Keble Road, Oxford OX1 3RH, UK}


\begin{abstract}
Starting with the well-known NFW dark matter halo distribution,
we construct a simple polytropic model 
for the intracluster medium
which is in good agreement with high
resolution numerical hydrodynamical simulations, apply this model to a very
large scale concordance dark matter simulation, and compare the
resulting global properties  with
recent observations of X-ray clusters, including the mass-temperature and
luminosity-temperature relations.
We make allowances for a non-negligible surface pressure,
removal of low entropy (short cooling time) gas,
energy injection due to feedback, and for a relativistic
(non-thermal) pressure component.
A polytropic index $n=5$ ($\Gamma=1.2$) provides a good
approximation to the internal gas structure of
massive clusters (except in the very central
regions where cooling becomes important), and allows
one to recover the observed $M_{500}-T$, $L_x-T$ and 
$T/n_e^{2/3}\propto T^{0.65}$  relations.
Using these concepts and generalizing this
method so that it can be applied to 
fully three-dimensional N-body simulations, one
can predict the global X-ray and SZE trends for any specified
cosmological model.
We find a good fit to observations when assuming that twelve 
percent of the initial baryonic mass condenses into stars,
the fraction of rest mass of this condensed component
transferred back to the remaining gas (feedback) is 
$3.9\times10^{-5}$, and the fraction of total pressure from
a nonthermal component is near ten percent.
\end{abstract}



\section{Introduction}   \label{sec:intro}
Gas in clusters of galaxies can be observed to large cosmological
distances by a variety of techniques, from X-rays (Bremsstrahlung) to
radio (S-Z effect).  But to utilize these observations it is necessary
to have a model for the state of the gas in clusters that is (a) motivated
by sound physical reasons, (b) able to fit
the observational data and (c) is
simple enough (i.e., {\it not} a numerical simulation!) to be applied broadly.
In this paper we attempt to present
such a model for the gaseous component in clusters of galaxies 
in order to provide predictions for global properties such as
temperature and X-ray luminosity.

For the dark matter (DM), the widely utilized
\citet{NFW}, or NFW, model satisfies all of the above criteria. 
Although we now 
know that it is not universal in two senses--- large variance 
in the central concentrations 
\citep{AFKK99,Jing00,BKSSKKPD01,KKBP01,FKM04,TKGK04}
and trends in properties with time and halo mass 
\citep{WBPKD02,Ricotti03,ZJMB03,WOB05,SMS05}---
it remains an extremely useful first basis for
analyzing and summarizing the properties of dark matter distributions.
We know, however that the gas in clusters does {\textit{not}} follow the dark
matter profile.  The well established central 
density profiles for dark matter
halos are roughly power laws: density $\rho$ depends on radius
$r$ as  $\rho \propto r^{-\alpha}$,
with $\alpha$ typically 1.0 (the NFW value) but ranging
from 0.5 to 1.5, depending on circumstances.  But the gas profiles show a
definite core 
\citep[$\alpha \rightarrow 0$; for a recent review see][]{Voit04}
and, as we shall see, one would overestimate the X-ray luminosity by
a large factor if one were to use the NFW or a steeper profile.

The construction of a satisfactory model will be guided by 
a few observed properties.
First, the gas is essentially a trace component
(approximately 1/7 of
total mass) and resides, in close to hydrostatic equilibrium, in a
potential which is well represented by NFW 
\citep[or its variants-- cf.][]{Zhao96}.
Furthermore, we know that there are efficient means of redistributing
energy/entropy within the cluster gas via, for example, 
turbulence \citep{KN03b} induced by merger shocks \citep{BN98} and
galaxy wakes \citep{SAP99,Sakelliou00}.
Additionally, other processes such as
conduction \citep{KN03a,DJSBR04},  
cosmic ray transport, and magneto-sonic wave transport \citep{Cen05}
may also be operating.
Also, appropriate boundary conditions
are required, since in both simple analytic models \citep[eg.][]{Bert85}
and detailed numerical simulations \citep[eg.][]{BN98,Frenk99}
the hydrostatic portion of the cluster gas is
terminated at an outer shock where the pressure is balanced by the
momentum flux of the infalling gas.  

A number of steps have already been made toward constructing
such a model.
\citet{MSS98} derived an analytic expression for a gas
distribution in hydrostatic equilibrium with an NFW
potential, assuming isothermality.  A more general 
expression for a polytropic equation of state was derived
by \citet{SSM98}; a similar functional form has been 
compared in detail with hydrodynamic simulations by \citet{AYMG03}.

In the simplest of such models the source of the gas heating is
gravitational, i.e. the gas energy comes from the same collapse
and virialization processes which determine the dark matter 
profile;  thus the energy per unit mass in the gas should be
approximately the same as in the dark matter.  This leads to an
expected self-similar scalings of mass $M$ and
luminosity $L$ with temperature $T$ 
of $M\propto T^{3/2}$ and $L\propto T^2$ \citep{Kaiser86}.
However, this expectation is in
contradiction with the observed relation, which is steeper
\citep{EdgeStew91,Markev98}.
\citet{Kaiser91} proposed that non-gravitational energy injection 
could lead to the observed relations.  This possibility has
been explored in the type of analytic model described here
using an NFW profile for the dark matter
\citep{SSM98,WFN00,SKSS04,LCM05,ALS05,SMGS05}, 
and with other profiles \citep{BBP99,BBLP02}.
The breaking of self-similarity can also be cast as
the modification of the initial gas entropy by thermal
and nonthermal processes, as explored in NFW-like potentials by
\citet{TozNor01}, 
\citet{KomSel01}, \citet{VBBB02}, and \citet{DosSD02};
on the other hand \citet{RoyNath03} argue that the entropy imparted to
the gas from gravitational processes alone is larger than previously
thought. 
Another impact on the gas energy comes from the fact that
approximately one tenth of the baryons in a typical
cluster are now in stellar form.  So one must allow for both
removal of the mass of this gas and of the associated energy
(or entropy) of this gas \citep{VB01,TozNor01,VBBB02,ScanOh04,BV05}.
Since the removed gas had short cooling times, low entropy, and
low total energy, the mean energy per unit mass of the remaining
gas is higher than before star and galaxy formation.

An issue not dealt with in these studies is non-thermal
sources of pressure.
Turbulence may provide in excess of 10\% of the total pressure
in Coma \citep{SFMBB04}; similar amounts of turbulent support
have been seen in simulations \citep{NorBry99,FKNG04}.
Clusters should also contain a population of relativistic 
particles arising from shocks, as recently reviewed by
\citet{Miniati04}, \citet{Sarazin04}, and \citet{Bykov05}.
Magnetic fields may also be dynamically important
\citep{CarTay02,EVP05,Bykov05}.

The basic goal of this paper is to start with a population
of dark matter halos from an N-body simulation, for which the DM
density profiles can easily be measured, and deduce the global properties
of the hot baryonic component in a physically well-motivated manner.
The ideal method should be as simple as possible while including
all the relevant components:  hydrostatic equilibrium inside a
dark matter halo potential;  gas energy per unit mass similar to
that of the dark matter, but modified by 
removal of low entropy gas and by feedback; 
appropriate outer boundary conditions; and pressure support
from a non-thermal component.
Some other processes necessary for detailed
models will not be included
because they are not in general required for obtaining global properties;
though
the results obtained here may need to be modified for those clusters
having distinctly cooler cores \citep{AllFab98}.
Finally, we will drop the limitation of a spherical NFW model
and generalize to any case for which the dark matter potential
is known.  While several of the papers quoted above have allowed
for some of these effects, none has included all in a fashion
that can be adapted to an arbitrary gravitational potential.

The next section reviews properties of the NFW model;
\S \ref{sec:igasep} derives the properties of an initially parallel
gaseous component and
\S \ref{sec:poly} derives these properties after the gas
rearranges itself;  \S \ref{sec:constr} presents 
the resulting profiles
and compares global properties with observed clusters.
All of these sections assume spherical symmetry.  In
\S \ref{sec:3d} we generalize the polytropic model
in order to remove geometrical constraints, concluding our
discussion in \S \ref{sec:conclusion}.

\section{The NFW Profile}   \label{sec:nfw}
\citet{NFW} have proposed a universal profile for dark matter halos,
which we will first review here to establish nomenclature.
Formally, the NFW profile extends to infinity and has logarithmically
diverging mass;
we will instead truncate the profile at the virial radius.
In this section we first
establish the properties of a distribution of matter with a
truncated NFW profile.  Assume that the density $\rho$ 
depends on radius $r$ as
\begin{equation} \label{eqn:rho}
\rho(r) = \left\{ \begin{array}{ll} 
\frac{\rho_1r_1^3}{r(r+r_1)^2} & r\leq r_v   \hspace{1cm} ,\\
0                              & r>r_v
\hspace{1cm} .
\end{array} \right.
\end{equation}
The virial radius $r_v$ is the radius within which the mean density
is 200 times the critical density $\rho_c$.
The parameters $C \equiv r_v/r_1$ and $\rho_1$ define the model.
It follows that
the mass $M$ is distributed inside $r_v$ as:
\begin{mathletters}
\begin{eqnarray} \label{eqn:mass}
M(r)&=& 4 \pi\  \rho_1 r_1^3\  g(x)    \hspace{1cm}, \\
g(x)&\equiv& \ln(1+x)-\frac{x}{(1+x)}  \hspace{1cm},
\end{eqnarray} 
\end{mathletters}
where $x\equiv r/r_1$.  Thus
$M_{tot} \equiv M(r_v) =  4\pi \ \rho_1 r_1^3\ g(C)$
is the total mass. 
The rotation curve,
$V_c^2(r)=GM(r)/r=4 \pi G \rho_1 r_1^2 g(x)/x$
provides a 
useful label for the mass distribution; 
it has a maximum at $x_{c,max}\approx 2.163$,
so the maximum circular velocity is
\begin{mathletters}
\begin{eqnarray} \label{eqn:circvmax}
V_{c,max}^2&=&4 \pi G \rho_1 r_1^2 G_{max} \hspace{1cm}, \\
G_{max}&\equiv& g(x_{c,max})/x_{c,max}\approx 0.2162
\hspace{1cm} .
\end{eqnarray} 
\end{mathletters}
Note that a given $M_{tot}$ and $V_{c,max}$ can be used
to completely define the matter distribution 
(as an alternative to $C$ and $\rho_1$).

The gravitational potential from this mass distribution is 
\begin{equation}  \label{eqn:phi}
\Phi(x) = -\frac{V_{c,max}^2 }{G_{max}}f(x) 
\hspace{1cm} ,
\end{equation} 
where
\begin{equation}
f(x) = \left\{ \begin{array}{ll} 
 \frac{\ln (1+x)}{x} -\frac{1}{1+C}  & x\leq C \\
 \left[ \frac{\ln (1+C)}{C} -\frac{1}{1+C} \right]\frac{C}{x}  & x>C \\
\end{array} \hspace{1cm} \right.
\end{equation} 
The total gravitational energy is then given by
\begin{mathletters}
\begin{equation} \label{eqn:W}
W_0 = \frac{1}{2}\int_0^{M_{tot}}{\Phi(r)dM}
 =  - G_{max}^{-1}V_{c,max}^2 M_{tot} H(C) \hspace{1cm} ,
\end{equation}
where
\begin{equation} 
H(C)\equiv \left[- \frac{\ln (1+C)}{(1+C)}+ 
\frac{C(1+C/2)}{(1+C)^2} \right] \frac{1}{g(C)} \hspace{1cm} .
\end{equation}
\end{mathletters}

Assuming velocities are isotropic for the bulk of the
matter distribution,
the velocity dispersion of the dark matter
in 1-D, $\sigma^2(r)$, obeys the equation
$\frac{d}{dr}(\rho \sigma^2)=-\rho\frac{d}{dr}\Phi$,
which has the solution
\begin{mathletters}
\begin{equation} \label{eqn:vdisp}
\sigma^2= \frac{V_{c,max}^2 }{G_{max}} S_C(x)x(1+x)^2  \hspace{1cm} ,
\end{equation}
where
\begin{eqnarray} \label{eqn:sofx}
S_C(x)= S_C(C) - \int_x^C{\frac{x' - (1+x')\ln (1+x')}{x'^3(1+x')^3}}dx'
\hspace{1cm} ,
\end{eqnarray} 
\end{mathletters}
with $S_C(C)$ a positive constant,
the value of which will be determined below.
The corresponding pressure is simply
$P=\rho \sigma^2 = 4 \pi G \rho_1^2 r_1^2 S_C(x)$,
so the average pressure is
\begin{mathletters}
\begin{equation} \label{eqn:pdef}
\overline{P} = \frac{1}{\frac{4}{3}\pi r_v^3} \int_0^{r_v}{ P d {\bf r}}
= {12 \pi G \rho_1^2 r_1^2 C^{-3}}\int_0^C{S_C(x)x^2 dx}  \hspace{1cm} ,
\end{equation}
and the surface pressure is
\begin{equation} \label{eqn:psurf}
P_s = 4 \pi G \rho_1^2 r_1^2 S_C(C) \hspace{1cm} .
\end{equation}
\end{mathletters}
A boundary surface pressure is required because of the jump in density and 
pressure at $r=r_v$, and would be provided in any realistic
physical model by the momentum flux from infalling
matter at the boundary.
One way to quantify
the pressure term represented by $S_C(C)$ is to estimate
the momentum per unit area transported
in by infalling matter.  If the rate of mass accretion
is $\dot{M}_{tot}$ and the accreted mass
(starting at rest from the turnaround radius)
is moving at freefall velocity $v^2_{ff}=V^2_c(r_v)$, then
$ P_s = \dot{M}_{tot} v_{ff} / (4 \pi r_v^2)$.
In the self-similar solution of \citet{Bert85}
$M_{tot} \propto t^{2/3}$. This should be approximately correct,
so we will take 
$\dot{M}_{tot} = 2qM_{tot}/(3t)$,
with deviations allowed for with use of the correction
factor $q$.
Clusters of galaxies are by definition in overdense regions,
so the surroundings will always appear to be close to
the critical density $\rho_c$, hence an appropriate time is 
$t = (6\pi G\rho_c)^{-1/2}$; 
thus we will adopt
\begin{equation}  \label{eqn:surfpres}
P_s = \frac{\sqrt{2}}{3}qV^2_c(r_v) \rho_s 
\left( \frac{\rho_c}{\rho_s} \frac{\bar{\rho}}{\rho_s} \right)^{1/2}
\hspace{1cm} ,
\end{equation}
where $\rho_s$ is the density at $r_v$, 
and $\bar{\rho}$ is the mean density inside $r_v$.
Combining this with Eqn. (\ref{eqn:psurf}),
it follows that
\begin{equation} \label{eqn:scofc}
S_C(C) = \left( \frac{2}{3}q^2 \frac{g^3(C)}{C^5} \frac{\rho_c}{\rho_1}
\right)^{1/2}
=\sqrt{2}q\frac{g^2(C)}{C^4}\left(\frac{\bar{\rho}}{\rho_c}\right)^{-1/2}
=q\frac{g^2(C)}{10C^4}
\hspace{1cm} .
\end{equation}

Alternatively, one may take the expression for the radial 
velocity dispersion
in an NFW halo derived by \citet{LM01} and evaluate it at
the virial radius $x=C$.  \citet{LM01} solved the Jeans equation
for an NFW density profile assuming a constant velocity 
anisotropy;  for the simplest case of isotropic orbits,
their Eqn. (14) yields
\begin{equation} \label{eqn:lmscofc}
\begin{array}{c} S_C(C) = 
\frac{\pi^2}{2} - \frac{{\rm ln}C}{2} - \frac{1}{2C} -
\frac{1}{2(1+C)^2} - \frac{3}{1+C} + 
\left( \frac{1}{2} + \frac{1}{2C^2} - \frac{2}{C} - \frac{1}{1+C} \right)
{\rm ln}(1+C) +  \\
\frac{3}{2}{\rm ln}^2(1+C) + 3{\rm Li}_2(-C) 
\end{array}
\end{equation}
with the dilogarithm Li$_2(x)=\int_x^0{\rm ln} (1-y)d{\rm ln} y$.
For the low concentration halos 
considered here, this gives pressures similar to Eqn. (\ref{eqn:scofc})
with $q\approx 4$.  We will use Eqn. (\ref{eqn:lmscofc}) in the
following, but have found simply using $q=4$ gives very similar results.
An intriguing feature of 
the NFW model is that the coarse-grained phase space density
$\rho(x)/\sigma^3(x)$ is a power law in radius, following $x^{-15/8}$
\citep{TayNav01,WABBD04}.  Using Eqn. (\ref{eqn:lmscofc}) to set
the surface pressure, the phase space density found by combining
Eqns. (\ref{eqn:rho}) and (\ref{eqn:vdisp})
does display in this power law behavior, with the appropriate slope.  
If we instead use (for example) $q=1$, then the same power law
still holds inside $0.5r_v$; only nearer the surface is there
a significant deviation.   This insensitivity to the exact
choice of surface pressure
helps explain why the approximation $q\approx 4$ works well.

\section{Initial Gas Energy and Pressure} \label{sec:igasep}
The goal of this paper is
to populate with gas, in a physically plausible fashion,
the potential well created by a dark matter
halo with known mass, radius, concentration, and 
maximum circular velocity.
Let the ratio of initial gas mass to total mass equal
the cosmic average, i.e. the total gas mass 
$M_{g,i} = \Omega_b M_{tot}/\Omega_m$.
We begin by assuming that initially the two components have a
parallel distribution--- which is what would be expected in the
absence of energy transport mechanisms.  Thus
the gas pressure is by hypothesis $P_g=(M_{g,i}/M_{tot})P$.

Now we apply the Virial Theorem to the whole, with allowance
for the non-negligible surface pressure:
$W_0 +2T_0-4\pi r_v^3 P_s = 0$.
Since the kinetic energy $T_0$ is related to the mean
pressure (in the absence of significant bulk motions) by
$2T_0=4\pi r_v^3 \overline{P}$,
we have, using Eqn. (\ref{eqn:pdef}), 
\begin{equation}
W_0 = -4\pi r_v^3 \left(\overline{P}-P_s \right)
= -3\langle \sigma^2 \rangle M_{tot}(1-\delta_s)
\hspace{1cm} ,
\end{equation}
where $\delta_s\equiv P_s/\bar{P}$ is the surface pressure
in terms of the mean.
When combined with Eqn. (\ref{eqn:W}),
this becomes
\begin{equation} \label{eqn:sigma}
3 \langle \sigma^2 \rangle (1 - \delta_s) = G_{max}^{-1} V_{c,max}^2 H(C)
\hspace{1cm} .
\end{equation}
Note this is taken from the definition of 
$\langle \sigma^2 \rangle 
\equiv \int{\sigma^2 dM}/M_{tot}
= (\frac{4}{3}\pi r_v^3 \overline{P})/M_{tot}$.
Rewriting Eqn. (\ref{eqn:sigma}) using
Eqns. (\ref{eqn:W}) and (\ref{eqn:vdisp}) gives
the dimensionless form of the Virial Theorem:
\begin{equation} \label{eqn:sk}
3(1-\delta_s)\int_0^C{S_C(x)x^2 dx} = H(C) g(C)
\hspace{1cm} .
\end{equation}
It follows that
\begin{equation}  \label{eqn:delta}
\delta_s = \frac{S_C(C)}{S_C(C) + C^{-3}H(C)g(C)}
\hspace{1cm} .
\end{equation}
Given $S_C(C)$, 
$\delta_s$ is known, and Eqn. (\ref{eqn:sigma}) 
can be rewritten as 
$\langle \sigma^2 \rangle = K(C) V_{c,max}^2$, where
\begin{eqnarray} \label{eqn:sigc} 
K(C)=\frac{1}{3}H(C)\frac{G_{max}^{-1}}{(1-\delta_s)} \hspace{1cm} ,
\end{eqnarray}
and the potential energy from Eqn. (\ref{eqn:W}) is now
\begin{equation} \label{eqn:WofC}
W_0  =  -3 K(C) (1-\delta_s) M_{tot} V_{c,max}^2
\hspace{1cm} .
\end{equation}
Thus the kinetic and potential energies can be specified, 
once the NFW parameters
for a dark matter halo
have been specified.

Earlier we
assumed a monatomic gas component, initially distributed
in the same manner as the dark matter.
What is the total energy of the gas?
Treating it as a tracer of negligible mass,
that is, assuming the gravitational potential is
totally determined by the dark matter,
$E_g=\onehalf M_{g,i} \langle 3 \sigma ^2 \rangle + 2\left(
{M_{g,i}}/{M_{tot}}\right)W_0$.
Combining this with Eqn. (\ref{eqn:WofC}) gives
\begin{equation} \label{eqn:egasjnos}
E_g=-\frac{3}{2}M_{g,i}V_{c,max}^2 K(C) (3-4\delta_s) 
\hspace{1cm} .
\end{equation}
Also, the gas surface pressure from Eqns. (\ref{eqn:psurf}) 
and (\ref{eqn:circvmax}) is
\begin{equation} \label{eqn:psurfC}
P_{s,gas} = \frac{M_{g,i}V_{c,max}^2}{4\pi r_1^3}
\frac{S_C(C)}{G_{max}g(C)}
\hspace{1cm} .
\end{equation}
Thus both the gas energy and surface pressure
are now known in terms of the
dark matter halo parameters.

\subsection{Allowance for Stellar Mass Dropout}   \label{sec:dropout}
Star formation will change the amount of energy in the
remaining gas, because that portion of the the gas which collapses
and is removed will have a lower entropy and a shorter cooling time
than is typical.  In our
idealized halo, this is the gas which would end up in the central
region, so we will estimate the change in energy by removing 
a fraction of the core; this removed fraction 
(corresponding to the mass in stars)
has the shortest cooling time
and lowest entropy.  
\citet{FHP98} estimated that
for clusters the stellar mass inside the virial radius
is roughly 0.19$h^{0.5}$ times the
gas mass, and \citet{SanPon03} find a median stellar to gas
ratio of 0.21; but lower values have been found 
by \citet{BPBK01} and \citet{LMS03}.  We will adopt
a stellar to gas mass ratio of 0.12 independent of cluster
mass, in reasonable agreement with the results of the latter two 
papers, assuming a LCDM model with $h$=0.7.
For simplicity we will assume
enough gas is turned into stars for this ratio to hold
generally.
In other words, the ratio of the mass in stars to
the final gas mass is $f_s=0.12$, with
$M_g=M_{g,i}/(1+f_s)$ remaining in the gaseous state.  This
will be done by removing all gas inside a radius $x_s$,
found by solving
$g(x_s)=g(C)f_s/(1+f_s)$.
The initial energy in this remaining gas is found
by integrating the previous expressions for the energy 
over only the mass $M_g$
outside of $x_s$:
\begin{equation} \label{eqn:egasj}
E_g=-\frac{3}{2}(1+f_s)M_g V_{c,max}^2 \left[ K(C) (3-4\delta_s) + 
 K_s(x_s)/G_{max} \right]
\end{equation}
with
\begin{equation} 
K_s(x_s) = \frac{1}{g(C)} \left[ \int_0^{x_s}S_C(x)x^2 dx - 
 \frac{2}{3}\int_0^{x_s}\frac{f(x)x dx}{(1+x)^2} \right]
\end{equation}
which, while unfortunately not as simple as before,
can still be determined once the dark matter halo parameters
are given.
By removing this gas we are both lowering the gas mass
and, more critically, increasing the energy per unit
mass of the remaining gas \citep{VB01}.

\subsection{Other Changes to the Gas Energy}   \label{sec:otherc}

The gas energy will change due
to the the work done by any
increase or decrease of the gas volume.  To calculate this
latter term, we will assume that the surface pressure does not 
change with radius, i.e. it is always given by Eqn. (\ref{eqn:psurfC}).
Let $C_f$ be the final outer radius of the gas distribution
in units of $r_1$; then
\begin{equation} \label{eqn:delep}
\Delta E_P=-\frac{4\pi}{3}r^3_1\left( C^3_f-C^3 \right)P_{s,gas}(C) 
=-\frac{1}{3}(1+f_s)M_gV_{c,max}^2\left( C^3_f-C^3 \right)
\frac{S_C(C)}{G_{max}g(C)} 
\hspace{1cm} .
\end{equation} 

In addition, we can expect there to
be energy input to the gas from feedback processes.  These are
primarily of two kinds:  wind and 
supernova shock energy deposited in the hot
gas, and heating due to output from accreting massive black holes in the
centers of the massive galaxies \citep{ScanOh04}.  
Dynamical friction on galaxies moving through the gas at
somewhat trans-sonic speeds may also be a source of energy input
\citep{Ost99,ElZKK04,FKNG04}.
To a good first approximation all of these effects
are proportional to the
gas mass (there is no evidence that the efficiency with which
gas is transferred to stars varies strongly and systematically with 
$V^2_{c,max}$ for moderately rich clusters).  
Let $\epsilon$ be a measure of the efficiency with which
gas is heated by the condensed component;
the energy input can then be written as
\begin{equation}   \label{eqn:delef}
\Delta E_f = \epsilon f_s M_g c^2
\hspace{1cm} .
\end{equation} 
The value of $\epsilon$ from energy output of supernovae can
be estimated as the product of the fraction of mass turned
into stars ($f_s$=0.12), the number of supernovae per solar
mass expected from a Salpeter initial mass function
($0.007M_\odot^{-1}$), and the energy input per supernova 
($10^{51}$ erg); this gives $\epsilon=2.8\times 10^{-6}$ or
$\epsilon f_s=3\times 10^{-7}$.
While assuming perfect efficiency in transferring this
energy into the gas is unrealistic, we find that by itself
this amount of energy has little impact on the gas profile,
consistent with previous results \citep{BBP99,VS99,BBLBCF01,SKSS04}.
The energy input from AGN is more substantial.
For a galaxy hosting a supermassive black hole, the
ratio of black hole to stellar mass will be
$\approx$ 0.0013 \citep{KG00,MerFer01}, so the ratio of 
black hole to total gas mass will be roughly $1.6\times 10^{-4}$.
Observational constraints give
the efficiency with which energy is released from 
accreting black holes to be $\approx 0.10$ \citep{YuT02}; 
adopting a conversion efficiency to mechanical energy 
of 30\% \citep{InSa01} leads to an
efficiency $\epsilon f_s=4.7\times 10^{-6}$ or 
$\epsilon=3.9\times 10^{-5}$.
We will adopt this last number in the rest of the paper.
This is equivalent to an energy input of 2 keV per particle
or 4 keV per baryon.
Another way of looking at this is to divide  $\epsilon f_s M_gc^2$
by a Hubble time to compute a typical luminosity;  for a halo
with a total mass of a few times $10^{14}M_\odot$, this is
of order $10^{44}$erg/s.  In other words, the energy input
from black holes is the same magnitude as the observed energy
radiated in X-rays.

\section{Polytropic Rearrangement}   \label{sec:poly}
Now assume that the gas rearranges itself  (changing its
density profile) through unspecified
processes into a polytropic distribution with polytropic
index $n$ or adiabatic index $\Gamma=1+n^{-1}$.
The commonly addressed cases are ($\Gamma,n$) = ($1,\infty$)
for isothermality, and (5/3,3/2) or (4/3,3) for non-relativistic
or relativistic isentropic fluids respectively.
We know that there is turbulence \citep{KN03a} induced by
merger shocks \citep{BN98} and galaxy wakes \citep{SAP99,Sakelliou00}.
Other processes driving the rearrangement
could include, for example,
conduction \citep{KN03a,DJSBR04}
or wave transport of energy 
\citep[via gravity, Alfven, or magnetosonic waves, e.g.][]{Cen05}.
This rearrangement means that the outer gas radius could
be larger or smaller than $r_v$.

For central pressure $P_0$ and density $\rho_0$, 
a polytropic distribution requires that the gas pressure
$P'$ and density $\rho'$ after rearrangement are related by
\begin{equation}  \label{eqn:polytrop}
P' = P'_0\left( \rho'/\rho'_0 \right)^{(1+1/n)}
  = P'_0\left( \rho'/\rho'_0 \right)^\Gamma  
\hspace{1cm} ,
\end{equation}
and the central isothermal gas sound speed is defined as
$V^2_{s0}={P'_0}/{\rho'_0}$.
Note that $\Gamma$ is not in fact the actual ratio of
specific heats; we require only that the gas has arranged itself in
polytropic fashion, as in Eqn. (\ref{eqn:polytrop}).
Fig$.$\,\ref{fig:gbrvrho}, kindly provided by Greg Bryan,
shows results from a high resolution adiabatic AMR simulation
of a massive ($\sim 10^{15}h^{-1}M_\odot$) cluster.  The
pressure--density relation in this calculation closely fits
that of a $\Gamma =1.15$ polytrope, and lends
support for the proposal that turbulent mixing (the only
one of the processes listed above that was included in this computation)
can lead to a fairly tight polytropic relation.
SPH simulations by \citet{LBKQHW00} resulted in a pressure-density
relation well described by a polytropic equation of state
with $\Gamma \approx 1.2$;  similar results are reported 
by  \citet{AYMG03} and \citet{BMSDDMTTT04}.
This result holds for both adiabatic and radiative simulations
\citep[but see][]{KTJP04}, 
and agrees well with the effective $\Gamma$ derived from
observed clusters by \citet{FRB01}.
\citet{SMGS05} find that $\Gamma=1.2$ offers the best consistency
with the assumption that the specific energy of the hot gas
equals that of the dark matter.
Interestingly, the purely adiabatic, spherical, 
and self-similar collapse
solution of \citet{Bert85} was also polytropic with 
$\Gamma \approx 1.17$.

An additional contribution to the pressure may come
from a relativistic component;  such a component
could be created for example at shock fronts, converting
part of the gas energy into cosmic rays.  While the
relativistic portion of the gas will contain a negligible
fraction of the mass, it may contribute significantly to
the total gas energy and pressure.
We allow for
the fact that, in addition to the gas pressure $P'$
there may be a nonthermal component having pressure
$P_{rel} = \delta_{rel}P'$
with total pressure
\begin{eqnarray*}
P_{tot}& = & P' + P_{rel} = P' (1+\delta_{rel}) .
\end{eqnarray*}
It is not obvious if $\delta_{rel}$ is maximum in the center, where
there may be injection of a relativistic fluid by an AGN, or in the
outer parts of the cluster, where there may be injection of
relativistic particle energy in boundary shocks.  Thus, for simplicity
we will take $\delta_{rel}$ = constant.

Given these relations,
the equation of equilibrium for a spherically symmetric
distribution, $dP_{tot}/dr=-\rho d\Phi/dr$,
becomes 
\begin{eqnarray*}
(1+\delta_{rel})V^2_{s0}\frac{\rho'_0}{\rho'}
\frac{d}{dr}\left( \frac{\rho'}{\rho'_0} \right)^{(1+1/n)}
= -\frac{GM_{tot}(r)}{r^2}  \hspace{1cm} .
\end{eqnarray*}
Thus
\begin{eqnarray*}
\left( \frac{\rho'}{\rho'_0} \right)^{1/n} - 1 
= -\frac{ \left[\Phi(r)-\Phi(0)\right] }{ V^2_{s0}(1+n)(1+\delta_{rel}) } 
= -\frac{ \beta j(x) }{(1+n)(1+\delta_{rel})}
\hspace{1cm} .
\end{eqnarray*}
This last is from Eqn. (\ref{eqn:phi}) with
\begin{mathletters}
\begin{equation} \label{eqn:beta}
\beta \equiv \frac{4 \pi G \rho_1 r_1^2}{V^2_{s0}}=\frac{V_{c,max}^2}{V^2_{s0}}G_{max}^{-1}
\end{equation}
and
\begin{equation}
j(x)\equiv \left\{ \begin{array}{ll} 
1 - \frac{\ln(1+x)}{x}  & x\leq C \\
1-\frac{1}{1+C}-\left[\ln(1+C)-\frac{C}{1+C}\right]x^{-1} & x>C
\hspace{1cm} .
\end{array} \right.
\end{equation}
\end{mathletters}
Thus,
\begin{mathletters}
\begin{equation} \label{eqn:grhovr}
\rho'(r)=\rho'_0 \left[ 1-\frac{\beta j(x)}{(1+n)(1+\delta_{rel})}
   \right]^n = \rho'_0\theta^n
\hspace{1cm} ,
\end{equation}
where
\begin{equation}
\theta = 1-\frac{\beta j(x)}{(1+n)(1+\delta_{rel})} 
\end{equation}
\end{mathletters}
is the familiar polytropic variable defined by \citet{Chandra}.
Eqn. (\ref{eqn:grhovr})
was first derived by \citet{MSS98} for the isothermal case, 
and \citet{WFN00} more generally.
The final gas radius can be smaller or larger than the
initial value.  Denoting this
radius in units of $r_1$ as $C_f$,
the gas mass can be written
$M_g=4\pi r_1^3 \rho'_0 L$,
where
\begin{equation}
L = L(n,\beta,C,C_f)\equiv  \int_0^{C_f} {\theta^n  x^2 dx} 
\hspace{1cm} .
\end{equation}
The thermal component will contribute a factor of
$\frac{3}{2}\int P'd^3x$ to the kinetic energy,
and the relativistic component
$3\int \delta_{rel} P' d^3x$, so
the rearranged total gas energy is 
$E_g'=\frac{3}{2}(1+2\delta_{rel})\int_0^{M_g}{V_s^2 dM} 
+\int_0^{M_g}{\Phi_{tot}dM}$.
Defining two more integrals
\begin{mathletters}
\begin{eqnarray} \label{eqn:i2int}
I_2&=& I_2(n,\beta,C,C_f)\equiv  \int_0^{C_f}{f(x)\theta^n x^2 dx} \\
I_3&=& I_3(n,\beta,C,C_f)\equiv \int_0^{C_f}{\theta^{1+n} x^2 dx} 
\end{eqnarray}
\end{mathletters}
we now have
\begin{equation} \label{eqn:egeqn}
E_g'=-M_{g}V^2_{s0} \left[-\frac{3}{2}(1+2\delta_{rel})\frac{I_3}{L} + 
       \beta \frac{I_2}{L} 
       \right]
\end{equation}
as the final energy.

\section{Constraints on the final temperature}   \label{sec:constr}
Suppose we have a dark matter halo for which the relevant
properties--- $M_{tot}, r_v, C, V_{c,max}$, etc.--- are known.
From the previous section, the final distribution of the gas can
be determined as a function of the two unknowns $\beta$
and $C_f$;  thus to specify
the final gas temperature and density distribution
it remains only to constrain these two parameters.
The first constraint is from conservation of
energy:  the final gas energy will equal the initial
energy plus changes to due star formation, expansion
or contraction, and feedback; i.e.
$E_g+\Delta E_p+\Delta E_f =E_g'$.
Combining this with Eqns. (\ref{eqn:egasj}), (\ref{eqn:delep}),
(\ref{eqn:delef}), and (\ref{eqn:egeqn})
yields
\begin{equation} \label{eqn:eqen}
\begin{array}{cc} 
\frac{3}{2}(1+f_s) \left[  
G_{max}K(C)\left(3-4\delta_s\right) + K_s(x_s) \right] 
- G_{max}\epsilon f_s\frac{c^2}{V_{c,max}^2}  \\
+ \frac{1}{3}(1+f_s)\frac{S_C(C)}{g(C)}\left( C^3_f-C^3 \right) =
\frac{I_2}{L}
-\frac{3}{2}(1+2\delta_{rel})\frac{I_3}{\beta L}
\end{array}
\end{equation}
(keeping in mind that $L$, $I_2$, and $I_3$ are functions
of $\beta$ and $C_f$).
A second constraint comes from the fact that the surface
pressure of the gas must match the exterior pressure, which
we have fixed at $P_{s,gas}(C)$.  This gives
\begin{equation} \label{eqn:eqpres} 
(1+f_s)\frac{S_C(C)}{g(C)}\beta L = (1+\delta_{rel}) \left[ 1-
   \frac{\beta j(C_f)}{(1+n)(1+\delta_{rel})} \right]^{1+n} 
\hspace{1cm} .
\end{equation}

Thus, given $r_v$, $C$, and $V_{c,max}$
for the dark matter halo, and appropriate choices
for $\Gamma$, $f_s$, $\epsilon$, and $\delta_{rel}$,
Eqns. (\ref{eqn:eqen}) and
(\ref{eqn:eqpres}) can be solved for $\beta$ and $C_f$, and
the central temperature
\begin{equation} 
kT_0 = \frac{\mu m_p}{G_{max}\beta}V^2_{c,max}
\end{equation}
is known, as is the density parameter $\rho'_0$,
so the gas distribution is fully specified.
The expected X-ray luminosity $L_X$ can then be calculated.
Following \citet{BBP99}, we will include both Bremsstrahlung
and recombination, which becomes important for temperatures
below 4 keV, by using the cooling function
\begin{equation} 
\Lambda(T) = 2.1\times 10^{-27} T^{1/2}
\left[ 1 + (1.3\times 10^6/T)^{3/2} \right]
\rm{ cm^3 erg\, s^{-1} . }
\end{equation}

\subsection{Simulated Halo Catalogue}   \label{sec:halocat}
The plausibility of this procedure can be evaluated by trying it
out on a population of many dark matter halos and comparing
the results to observed clusters.  The halos we use here come
from an N-body simulation designed to be in concordance with
observational constraints.
The simulation is of a periodic cube 1500$h^{-1}$Mpc on a side
containing N$=1260^3=2\times 10^9$ particles.   
The cosmology was chosen to be a standard LCDM power
law model with the following parameters:
baryon density $\Omega_b=0.047$; Cold Dark Matter density 
$\Omega_{\rm CDM}=0.223$ (hence total matter density 
$\Omega_m=0.27$);
cosmological constant $\Omega_\Lambda=0.73$ (thus spatially flat);
Hubble constant given by $h=H_0/(100{\rm km~s^{-1} Mpc^{-1}})=0.70$
(hence $\Omega_bh^2=0.02303$); primordial scalar spectral index
$n_s=0.96$; and linear matter power spectrum amplitude
$\sigma_8=0.84$.   These values are consistent within one 
standard deviation to
those derived either from WMAP data or from WMAP combined with 
smaller angular scale CMB experiments and galaxy data 
\citep{SpergWMAP}.
The initial conditions
were generated using the publicly available code GRAFIC2 \citep{Bert01}
to compute initial particle velocities and
displacements from a regular grid.  Since the memory required
to hold a $1260^3$ grid is 8 gigabytes, it was necessary to 
modify the single level portion of this program by adding
message-passing commands in order
to distribute the mesh among several
processors.

The simulation was carried out with the TPM (Tree-Particle-Mesh)
code \citep{BO03}, using 420 processors on the
Terascale Computing System at the Pittsburgh Supercomputing Center;
it took not quite five days of actual running time.
The box size and particle number determine
the particle mass of $1.26\times 10^{11}h^{-1} M_\odot$.
The cubic spline softening length was set to $17h^{-1}$kpc.
A standard friends-of-friends (FOF) halo finding routine was run
on the redshift $z=0$ box, using a linking length 
$b=0.2$ times the mean interparticle separation \citep{LC94};
this yielded 575,125 halos with both a FOF mass above
$2\times 10^{13}h^{-1} M_\odot$ and a virial mass above
$1.75\times 10^{13}h^{-1} M_\odot$.
The PM mesh used in TPM contained 1260$^3$ cells, and at redshift zero
all PM cells with an overdensity above 39 were being followed
at full resolution, so these objects had the full force 
resolution of TPM.  For the range of parameters used here,
clusters with $kT>2$keV contained more than 200 particles within $r_v$.

For each halo, the position of the most bound particle is
taken to be the cluster center.  Then $M_{tot}$ and $r_v$
are measured, as are $V_{c,max}$ and the radius of
maximum circular velocity $r_{c,max}$;  this latter gives
the concentration $C=2.163r_v/r_{c,max}$.  
This defines the equivalent NFW model halo, i.e. that NFW
model closest to the computed dark matter halo.
With this 
information, the procedure outlined above can be carried
out on each halo to compute the gas density and temperature.

\subsection{Resulting Profiles}   \label{sec:resprof}
\begin{figure}
\plotone{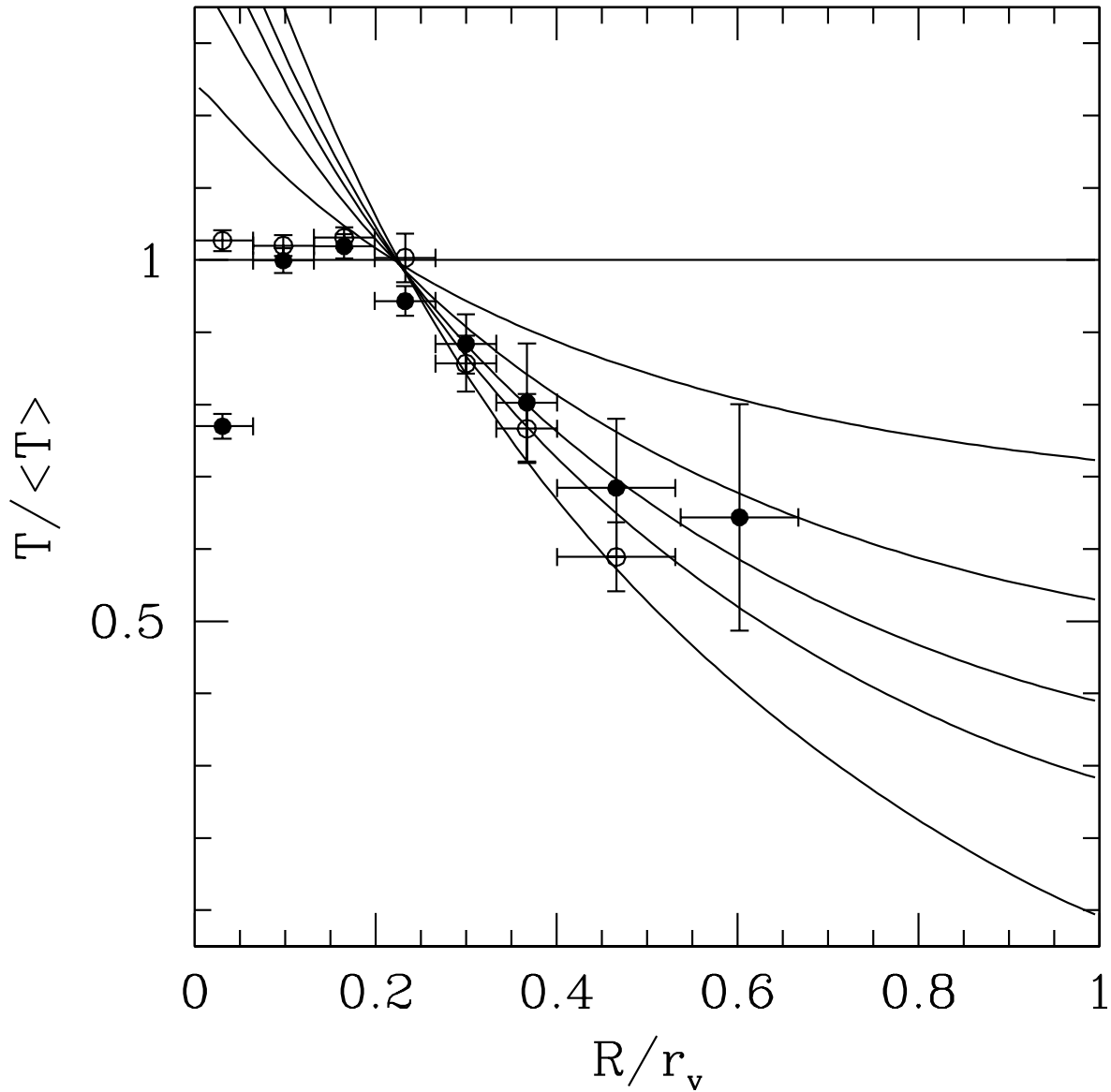}
\caption{Projected
temperature profile for $\Gamma=$1, 1.1, 1.2, 1.3, 1.4, and 1.67
(solid lines, from top to bottom at outer radius).  $<T>$ is
calculated by evenly weighting all radii inside $r_{v}/2$.
The points are from \citet{DeGM02}; filled and
open circles are clusters with and without cooling flows.
The parameters used to calculate the profile are $C=4, V_{c,max}=1200$
km/s, $M_{tot}=4\times10^{14}h^{-1}M_\odot$, and
$r_{v}=1.2h^{-1}$Mpc.  Aside from the innermost region, the model
with a $\Gamma$ of 1.2-1.4 satisfactorily represents the $T$ falloff
in the outer regions, which contain most of the gas mass.
\label{fig:tprof} }
\end{figure}

Given this set of halos, it remains to specify
$\Gamma$, $f_s$, $\epsilon$, and $\delta_{rel}$.
Let us first consider the appropriate $\Gamma$.
Fig$.$\,\ref{fig:tprof} shows the projected temperature profile
for different values of $\Gamma$;  this profile
was computed by integrating
the  emission-weighted temperature  along the line
of sight.  To normalize the curves, the mean temperature 
$\langle T\rangle$ was
calculated by evenly weighting all radii inside $r_{v}/2$;
this was done to correspond with the method of \citet{DeGM02},
who measured the mean profile for clusters 
with and without cooling flows--- shown in the Figure as
filled and open circles, respectively. 
Examination of Fig$.$\,\ref{fig:tprof} shows that
$\Gamma$=1.2--1.4, corresponding to polytropes with
index $n$=2.5--5, provides adequate fits to
the outer parts of the clusters, within which
resides most of the gaseous mass
\citep[see also the discussion in][]{SMGS05}.
\citet{AYMG03} have shown that a $\Gamma=1.18$ model is
a good fit to the average temperature profile measured by 
\citet{MFSV98}; this latter measurement has been
confirmed by \citet{DeGM02}, \citet{PJKT05}, and \citet{VMMJFvS05}.
As discussed above (\S \ref{sec:poly}), $\Gamma=1.2$ is also a good fit to
hydrodynamical simulations
\citep{LBKQHW00,LNNBBM02,AYMG03,BMSDDMTTT04,KTJP04}.
The lack of an isothermal core will not lead to a serious 
overestimation of
luminosity or emission-weighted temperature because the
volume of this central region is small.
This was tested by taking the $\Gamma$=1.2 profile and
imposing an isothermal core,  matching the density and
pressure at $0.2r_v$;  the resulting changes in 
emission-weighted temperature and luminosity were generally
less than 10\%.
However, neglecting cooling will reduce the
scatter in the $M_{500}-T$ and $L_x-T$ relations \citep{MBBPH04}.

The effect of the polytropic rearrangement on the radial profile
of the gas can be seen in Fig$.$\,\ref{fig:profs}.
The example halo used here has physical parameters
$M_{tot}=4\times 10^{14}h^{-1} M_\odot$, $r_v=1.2 h^{-1}$Mpc,
$C=4$, and $V_{c,max}=1200$km/s.
The top two panels show the  temperature (relative
to $T_{ew}$, the mean emission-weighted  temperature inside
a radius $R_{500}$ containing an density of $500\rho_c$)
and density (relative to $200\rho_c$) as a function
of radius.  It is instructive to compare to the original
NFW distribution, shown as a dot-dashed line; in this case the
central temperature goes to zero, as the density profile
has a cusp.  The polytropic rearrangement (taking 
$\Gamma=1.2$ and $f_s=\epsilon=\delta_{rel}=0$, 
shown as a dotted line) increases the central temperature
while decreasing the density, removing the cusp
(with a correspondingly dramatic lowering of the
X-ray luminosity, as we shall see).
This is seen more clearly in the third panel, which shows the
ratio of gas to dark matter mass interior to a
given radius, in terms of the cosmic average:  inside
the dark matter core radius, the  gas fraction declines
sharply.
These temperature and density profiles result in the
``entropy'' profile shown in the final panel of
Fig$.$\,\ref{fig:profs}, taking the definition of
entropy to be $T\rho^{-2/3}$.
The polytropic profile has a slope close to $r^{1.1}$ near
the virial radius, and is shallower nearer the cluster core;
this behavior has in fact been observed in a wide range
of clusters \citep{PSF03,PA05,PJKT05}.  This behavior has
been derived before in analytic models assuming the gas
is shock heated \citep{TozNor01}, and is also seen in
hydrodynamic simulations \citep{LBKQHW00,BMSDDMTTT04,KTJP04}.

This change in profile has a strong impact on other observable
cluster properties, as is shown in 
Fig$.$\,\ref{fig:rels}.  In these plots the temperature
is taken to be the mean emission-weighted $T$ 
inside $R_{500}$, $T_{ew}$.  In clusters with more complicated
structure this measure may not coincide well with the
spectroscopically measured $T$, as pointed out by
\citet{MRMT04}, who provide an alternative measure.
However, for the simplified models here, the difference
between emission weighting and the \citet{MRMT04} 
spectroscopic-like measure is only a few percent at most.
The lines show the median value 
as a function of $T_{ew}$,
found using the dark
halo catalogue described in \S \ref{sec:halocat}.
The first impact of
the polytropic rearrangement is to increase the observed
temperatures.  This is clearly demonstrated in the left-hand
panel, which gives the mass-temperature relation.  
The points are from \citet{ReipBohr02}, as adjusted 
by \citet{MBBPH04};
here the mass is $M_{500}$,
the mass inside a sphere containing mean density 500$\rho_c$. 
The polytropic model distribution (dotted line) resembles that
assuming an NFW profile, only shifted to higher $T$;
note assuming an NFW gas profile leads to significant
disagreement with the observed relation, while switching
to a polytropic model provides much superior agreement.
This is also true in
the right-hand panel, which shows the bolometric X-ray
luminosity as a function of $T$; the data points
are the subset of the ASCA cluster catalogue 
\citep{Horner01} described in \citet{MBBPH04}.
The polytropic model,
without the central cusp, yields a lower luminosity
than the NFW profile.  The slopes of both the $L_x-T$ and
$M_{500}-T$ relations retain the same self-similar
values, however.

The next physical input is the fraction of gas 
which collapses into stars.  As discussed above,
this is roughly one eighth the 
the gas mass inside the virial radius, or $f_s=0.12$.
Since the stars in clusters are old, this fraction will
hold for all moderately low redshifts.
As shown as short-dashed lines in 
Fig$.$\,\ref{fig:profs}, assuming $f_s=0.12$
for a typical cluster (keeping $\Gamma=1.2$)
increases the temperature slightly and reduces the gas density.
Since the temperature change is not large, this has
little effect on the
$M_{500}-T$ relation.
However, gas removal for star formation,
which increases the mean energy per particle for the
remaining gas,
leads to lower densities and so has a significant
impact on the $L_x-T$ relation.  For the most massive
(hottest) clusters, the predicted luminosity is in fact
close to that observed;  however, the self-similar slope
of the relation is still preserved, so for less massive clusters
$L_x$ is overestimated.  

The next required physical input is the amount of energy
from feedback coming from supernovae and active galactic nuclei,
discussed in \S \ref{sec:otherc}.
The results of including feedback of $\epsilon = 3.9\times 10^{-5}$
in the $\Gamma=1.2, f_s=0.12$ model are shown as
long-dashed lines in
Figs$.$\,\ref{fig:profs} and \ref{fig:rels}.
As one would expect, the radial profile has a higher
temperature and lower density.  However, the effect
of feedback differs from those considered previously,
because the resulting relations are no longer self-similar.
For massive clusters with $V_{c,max}>1000$km/s
or $kT>10$keV, feedback is of little importance
because the added energy is small compared to 
the gravitational energy,
but for smaller masses it can have a significant
impact.  One can see a steeper slope in the
$M_{500}-T$ relation, but the most significant effect is on
the luminosity, which in shape now more closely
resembles the observed distribution.

The remaining physical effect left to include is nonthermal
pressure.  
We will take $\delta_{rel}=0.1$;  the nonthermal sources
of pressure may in fact contribute a few tens
of percent of the total \citep{Miniati04}.
The results can be seen by
comparing the solid ($\delta_{rel}=0.1$) and long-dashed
($\delta_{rel}=0$) lines in 
Figs$.$\,\ref{fig:profs} and \ref{fig:rels}.
With this additional support, less kinetic energy
is required at a given pressure. 
Thus, while the density profile is little changed,
the resulting gas distribution is somewhat cooler,
and the emission weighted temperature is lower at
a fixed $M_{500}$ or $L_x$.

\begin{figure}
\plotone{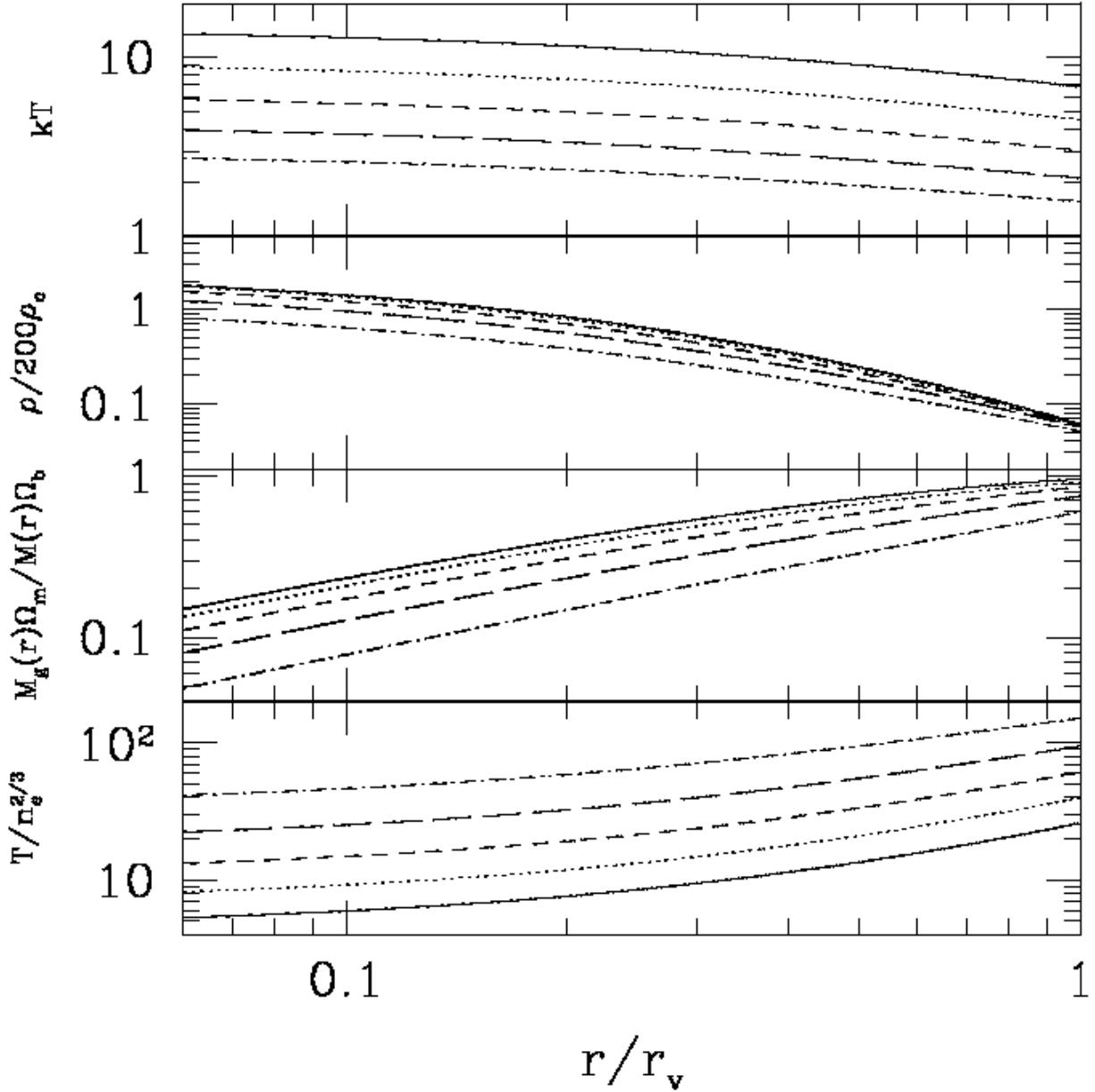}
\caption{Radial distributions for masses
$M_{tot} = 10^{15}, 5\times10^{14}, 2.5\times10^{14}, 1.25\times10^{14}$,
and $6.25\times10^{13} h^{-1}M_\odot$
(line types from top to bottom in uppermost panel).
Shown, from top, are temperature, density relative to 200
times critical, the gas mass fraction interior to $r$,
and the entropy $T/n_e^{2/3}$ 
(here $n_e$ is in units of cm$^{-3}$).
Parameters used are $\Gamma=1.2$, $f_s=0.12$, $\epsilon=3.9\times 10^{-5}$,
and $\delta_{rel}=0$.
\label{fig:mdprofs} }
\end{figure}

The departure from self-similar scaling is shown further in
Fig$.$\,\ref{fig:mdprofs}, which displays the radial 
profiles of temperature, gas density, gas fraction, 
and entropy for clusters of mass 
$M_{tot} = 10^{15}, 5\times10^{14}, 2.5\times10^{14}, 1.25\times10^{14}$,
and $6.25\times10^{13} h^{-1}M_\odot$.  Star formation and feedback were
included with $f_s=0.12$ and $\epsilon=3.9\times 10^{-5}$,
but not a relativistic component.  For the least massive cluster
we took $C=4$, $V_{c,max}=700$km/s, and $r_v=650h^{-1}$kpc,
scaling for the others as $C\propto M^{-0.13}$,
$V_{c,max}\propto M^{1/3}$, and $r_v\propto M^{1/3}$;
this is in reasonable agreement with our N-body cluster
catalog.  With these parameters, the emission weighted
temperatures inside $R_{500}$ are
$kT_{ew}$=10.1, 6.7, 4.5, 3.1, and 2.2 keV, respectively.  
For decreasing mass, the temperature and density profiles
become increasingly shallow, leading to a faster decrease in
X-ray luminosity.   
The ejection of gas following feedback energy injection
leads to a gas fraction (relative to the universal value)
less than unity at the virial radius;
with the full halo catalog and these parameters we
find for halos in the range $1-2\times10^{14}h^{-1}M_\odot$,
the gas fraction at the virial radius is $0.72\pm 0.09$
(one standard deviation).
This result is
in agreement with simulations including both heating and
cooling:  \citet{MTKP02} and \citet{KNV05} find that for
halos with $M \approx 10^{14}h^{-1}M_\odot$ the hot gas 
fraction inside a radius enclosing overdensity 
$\approx 100\rho_c$ is in the range 0.6--0.7, while
\citet{EBMMTDDSTT04} find slightly higher values of 0.7--0.8.
Observational estimates give similar values with a
higher scatter \citep{Evrard97,MME99,SanPon03}. Both
these observational and the theoretical studies suggest
that more massive clusters have higher hot gas fractions,
behavior which is reproduced here.

\begin{figure}
\plotone{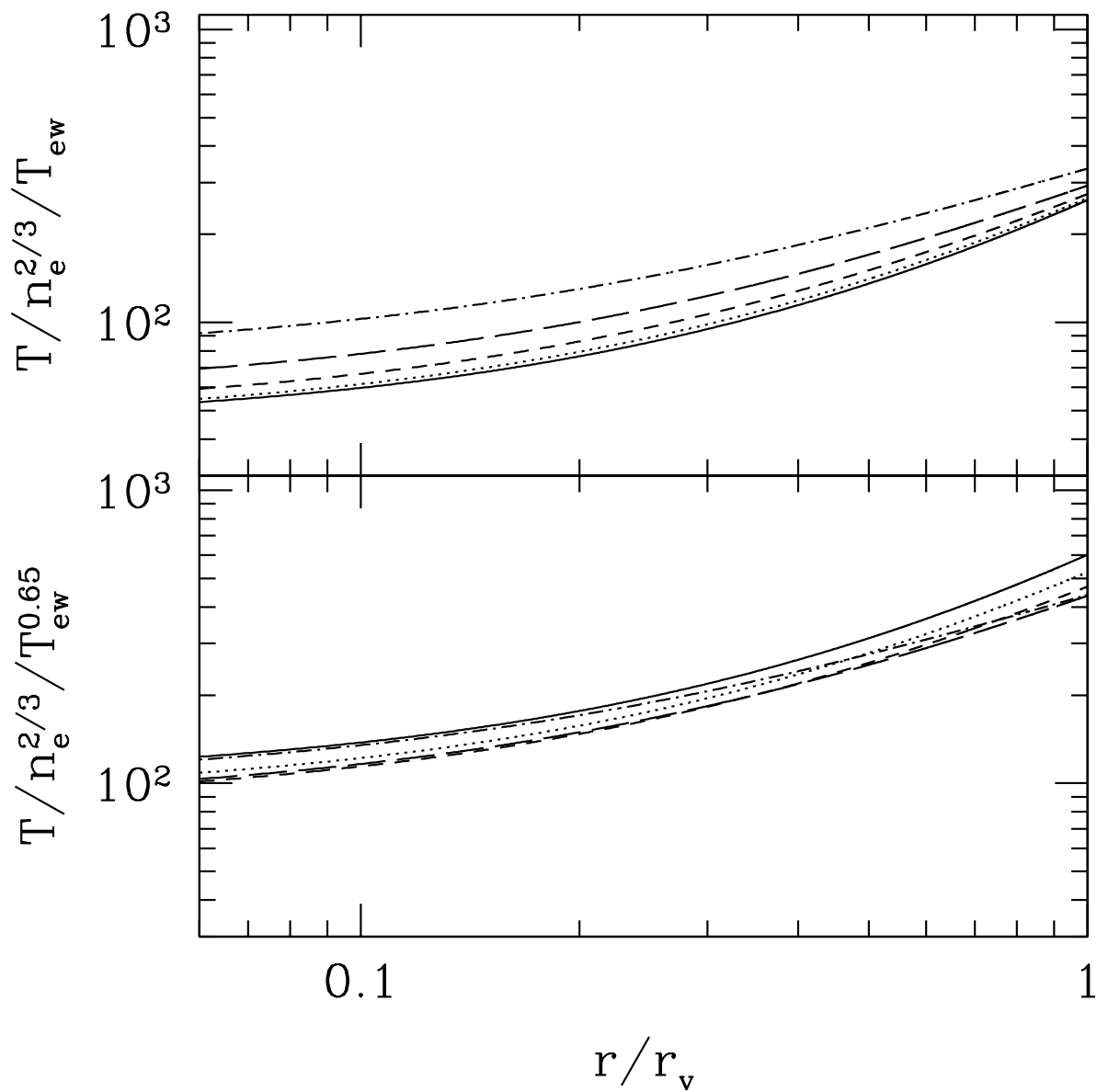}
\caption{Radial distribution of entropy $T/n_e^{2/3}$ for 
cluster masses
$M_{tot} = 10^{15}, 5\times10^{14}, 2.5\times10^{14}, 1.25\times10^{14}$,
and $6.25\times10^{13} h^{-1}M_\odot$
(line types as in Fig$.$\,\ref{fig:mdprofs})
scaled by $T_{ew}$ (top panel)
and by $T_{ew}^{0.65}$ (bottom panel).
\label{fig:mdents} }
\end{figure}

The bottom panel in
Fig$.$\,\ref{fig:mdprofs} displays the 
entropy $T/n_e^{2/3}$,
with $n_e$ the electron
density in units of cm$^{-3}$ (and $T$ in keV), which can be compared
directly with the observations listed above.
The entropy profiles have
a slope close to the observed value of $r^{1.1}$ at the 
virial radius,
but this slope quickly becomes shallower for smaller radii.
However,
we have not taken into account cooling, which would 
be more important near the center, steepening
the inner entropy profile \citep{MBBPH04}.
It has been observed that the 
entropy in clusters scales as $T^{0.65}$ 
\citep{PSF03,PA05,PJKT05},
rather than linearly in temperature as one would
expect for self-similar scaling.
Both of these scalings are shown in
Fig$.$\,\ref{fig:mdents}; the upper panel gives
$T/n_e^{2/3}/T_{ew}$  and the lower panel
$T/n_e^{2/3}/T_{ew}^{0.65}$.
It is clear from this Figure that
the observed  $T^{0.65}$ scaling is followed quite closely.
However, at radii near $r_v$ the observational picture
is unclear; with a sample of 14 nearby clusters,
\citet{Neumann05} found that the outer regions followed
self-similar scaling and may be affected by the accretion
of cooler material.
Also, for poorer systems than considered in this
paper ($kT<2$keV), we find that this scaling
breaks down; such groups may have different
entropy profiles than are seen in richer systems \citep{MFBGH05}.

\section{Generalization to an Arbitrary Dark
Matter Potential} \label{sec:3d}
The virtue of the method presented in this paper is not 
just in
the equations presented in the preceding sections, whose purpose
was to establish physical and mathematical principles and assess
the plausibility of the results, but also in its ability to efficiently
model complex asymmetrical systems containing substantial substructure.
In this section we relax the previous assumption of spherical
symmetry and apply the same method to more complex DM potentials.
Suppose that a cluster
potential is known from an accurate DM integration; the cluster
will likely be aspherical and contain significant substructure.
Then, if we are satisfied by the ability of an equilibrium polytrope
to model the gas in a cluster of galaxies, the
integration of the equation of equilibrium 
$\vec{\nabla}P_{tot}=-\rho\vec{\nabla}\phi$
gives
\begin{equation}
(1+\delta_{rel})\frac{\Gamma}{\Gamma-1}\frac{P(\vec{r})}{\rho(\vec{r})} = 
-\phi(\vec{r}) + 
(1+\delta_{rel})\frac{\Gamma}{\Gamma-1}\frac{P_0}{\rho_0} + \phi_0
\hspace{1cm} .
\end{equation} 
The last two terms comprise a constant of integration;
here $\phi_0$ is the potential minimum located at 
position $\vec{r}=\vec{r_0}$,
and the pressure and density at this point are designated by $P_0$
and $\rho_0$.
Then, making the definition
\begin{equation} \label{eqn:3dtheta}
\theta(\vec{r}) \equiv 1 + \frac{\Gamma-1}{(1+\delta_{rel})\Gamma}
  \frac{\rho_0}{P_0}\left(\phi_0-\phi(\vec{r})\right)
\hspace{1cm} ,
\end{equation} 
the pressure and density are simply
\begin{mathletters}
\begin{eqnarray} \label{eqn:p0rho0}
P(\vec{r}) = P_0 \theta(\vec{r})^\frac{\Gamma}{\Gamma-1} \\
\rho(\vec{r}) = \rho_0 \theta(\vec{r})^\frac{1}{\Gamma-1}
\end{eqnarray}
\end{mathletters}
where $\theta(\vec{r})$ is essentially the same polytropic
variable defined by \citet{Chandra}.  
Thus for an equilibrium polytropic gas residing in a known
potential $\phi(\vec{r})$, the determination of the structure
is reduced to the determination of the two numbers
$P_0$ and $\rho_0$. 
Adopting the approach taken in the previous
sections, these constants can be determined by satisfying two equations
of constraint on the final energy and the surface pressure.

We carried out this procedure on the same N-body simulation
used previously in the following manner.  A set of particles
identified as a cluster is placed in a nonperiodic 3-D grid.
The grid cell size $l$ is set to four
times the N-body particle spline softening length, as scales
smaller than this can be affected by numerical resolution 
issues; increasing or decreasing $l$ by a factor of two had
little impact on the results.  The dark matter density in each
cell $k$, $\rho_k$, is found from the particle positions using
cloud-in-cell (CIC), and the gravitational
potential $\phi_k$ on the mesh is calculated from the density using a 
nonperiodic FFT \citep{HE81}.  
The position of the cell with the minimum
potential $MIN(\phi_k)=\phi_0$ is 
taken to be the center of the cluster, $\vec{r_0}$.
The cluster velocity is estimated as the mean velocity of
the 125 particles closest to $\vec{r_0}$;  this mean is subtracted from
all the particle velocities. 
Then the DM kinetic energy per unit volume 
$t_k=\rho v^2$ is found in each cell: as with the mass, the kinetic
energy of each particle is distributed among 8 cells using CIC.
The virial radius $r_v$ and DM mass $M_{tot}$ are found from
the density distribution.  The $N_{cl}$ cells inside $r_v$ are
identified, as are the $N_{b}$ cells in a buffer region
of width $r_{buf}$
surrounding the cluster with centers in the range $r_v<r<r_v+r_{buf}$.
The buffer width was set at 9 cells, or $r_{buf}=153h^{-1}$kpc
for the simulation used here.  The gas surface pressure is taken
to be the mean value (assuming velocities are isotropic)
inside this buffer region:
\begin{equation} \label{eqn:3dpsurf}
P_s = N_{b}^{-1}\sum\limits_{k=1}^{N_{b}} 
\frac{1}{3} \frac{\Omega_b}{\Omega_m} t_k
\end{equation} 
where the sum is over all cells in the buffer region.
Assuming gas traces the DM,
the gas mass inside $r_v$ is originally
$M_{g,i}=\frac{\Omega_b}{\Omega_m}M_{tot}$.  As before,
the portion of this gas which is turned
into stars is $f_sM_{g,i}/(1+f_s)$.
To decide which portion of the gas becomes stars,
cells are ranked by binding energy $\rho_k\phi_k+\onehalf t_k$;
starting with the most bound cell, the initial $\rho_k$ and
$t_k$ are set to zero for each cell in turn until the gas mass removed,
$\frac{\Omega_b}{\Omega_m}\rho_k l^3$, totals to 
$f_sM_{g,i}/(1+f_s)$.  The original
gas mass inside $r_v$ is then 
\begin{equation} \label{eqn:3dmass}
M_g = \frac{M_{g,i}}{1+f_s} = 
\sum\limits_{k=1}^{N_{cl}} \frac{\Omega_b}{\Omega_m}\rho_k l^3 
\end{equation} 
the sum being over all cells inside $r_v$,
and the initial gas energy is
\begin{equation} \label{eqn:3dinite}
E_g = \sum\limits_{k=1}^{N_{cl}}  \frac{\Omega_b}{\Omega_m}
\left\{ \phi_k\rho_k+\onehalf t_k \right\} l^3
\hspace{1cm} .
\end{equation} 
As before, this energy can be supplemented by feedback energy
$\Delta E_f = \epsilon f_s M_g c^2$.

As in the 1-D case, this gas is assumed to rearrange itself
into a polytropic distribution with $\Gamma=1.2$.  It only
remains to specify $P_0$ and $\rho_0$, which are fixed
by the final energy and surface pressure.
For a given initial choice of ($P_0,\rho_0$), the final
gas density and pressure can be found 
after calculating $\theta_k$ for each cell from
Eqn. (\ref{eqn:3dtheta}).
As before, the initial energy may be changed by the inflow
or outflow of gas.
The final radius of the gas, $r_f$, is found by moving
outwards from the cluster center until mass $M_g$ is enclosed,
i.e.
\begin{equation} \label{eqn:3dfm}
M_g=\sum\limits_{k=1}^{N_{f}} \rho_0 \theta_k^\frac{1}{\Gamma-1}l^3
\hspace{1cm} ,
\end{equation} 
where the sum is over the $N_{f}$ cells inside $r_f$.
Similarly to Eqn. (\ref{eqn:delep}), we will assume that the
surface pressure does not change with radius, so the change
in energy due to expansion or contraction is proportional
to the change in volume.  This means 
$\Delta E_P = (4\pi/3)(r_v^3-r_f^3)P_s$,
with $P_s$ given by Eqn. (\ref{eqn:3dpsurf}).
Now we have all the information required for
the first constraint on ($P_0,\rho_0$), namely the conservation
of energy, where the 
final gas energy is 
\begin{equation} \label{eqn:3dfe}
E_f=\sum\limits_{k=1}^{N_{cl}} \left\{ 
\rho_0 \theta_k^\frac{1}{\Gamma-1} \phi_0
+ \frac{3}{2}P_0\theta_k^\frac{\Gamma}{\Gamma-1} \right\}l^3
= E_g + \Delta E_f + \Delta E_P
\hspace{1cm} .
\end{equation} 
The second constraint is the mean pressure in 
the $N_{b,f}$ buffer cells between $r_f$ and $r_f+r_{buf}$; 
this is assumed to match the original value:
\begin{equation} \label{eqn:3dfp}
N_{b,f}^{-1} \sum\limits_{k=1}^{N_{b,f}} P_0\theta_k^\frac{\Gamma}{\Gamma-1}
= P_s
\hspace{1cm} .
\end{equation} 
Thus after an initial estimate for ($P_0,\rho_0$), it is
now possible to iterate to a solution satisfying
Eqns. (\ref{eqn:3dfe}) and (\ref{eqn:3dfp}).
This solution provides the full three
dimensional pressure and density of the gas with allowance
for substructure, triaxiality, etc.

\begin{figure}
\plotone{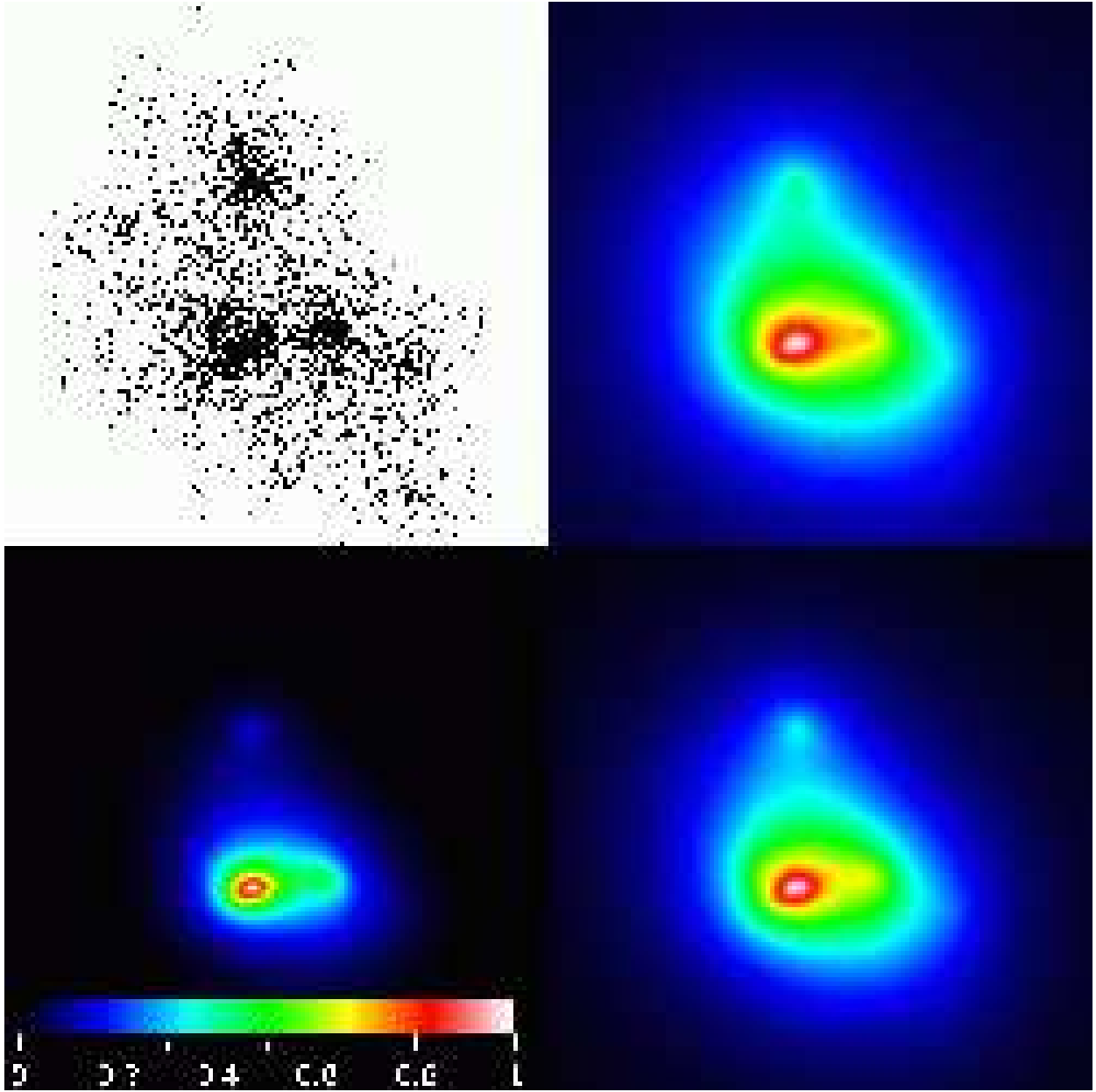}
\caption{  Adding gas to a dark matter halo 
(drawn from a larger simulation)
containing significant substructure. 
{\it Upper left:} DM particle distribution.
{\it Upper right:} projected gas surface density.
{\it Lower left:} X-ray luminosity.
{\it Lower right:} integral of the pressure along the line
of sight, proportional to the SZE decrement.
The color scale is normalized to the maximum value;
the volume shown is 6.4$h^{-1}$Mpc on a side.
\label{fig:3dexmpl} }
\end{figure}

An example of the resulting gas distribution 
for one halo taken from the catalog is given
in Fig$.$\,\ref{fig:3dexmpl}.  The simulation particle
positions are shown in the upper left-hand panel; 
there are several large substructures in the process 
of merging with the main objects.  The volume shown
is a cube 6.4$h^{-1}$Mpc on a side.  The upper right-hand
panel shows the projected gas surface density obtained
by the method just described.  With the gas density and
temperature, maps can be made for the X-ray emission and
SZE, as shown in the lower panels.
The scale is linear, with black a factor of 100 
below white.
The results of this procedure employed on the entire set of
halos used before are displayed in Fig$.$\,\ref{fig:3drels}.
A stellar fraction $f_s=0.12$ and feedback $\epsilon=3.9\times 10^{-5}$
were included.  For each plot, the median value is shown
as a solid line and the shaded region encloses 68\% of the
halos at that temperature.  Also shown as dashed lines are
the results of the method of \S \ref{sec:constr} 
based on the NFW profile;
the scatter seen using this latter method is somewhat smaller
than that shown for the full 3-D method, as the latter
realistically includes substructure and triaxiality.
We are neglecting cooling, which would increase the scatter
further, and tend to increase the luminosity at a given
temperature \citep{MBBPH04}.
Use of the NFW approximation appears to have little 
effect on either the $M_{500}-T$ or the $L_x-T$ relation, relative to
using the full particle distribution.
Examination of individual clusters shows that the spherically
averaged gas profiles resulting from the N-body potentials
are slightly shallower in both temperature and density,
although the amount of gas inside $R_{500}$ is the same for both
methods;  the larger clumping factor when using the true density,
with triaxiality and substructure, increases the luminosity
enough to compensate for this.

\section{Conclusion} \label{sec:conclusion}

The NFW model has provided a useful description of the
distribution of matter inside collisionless DM halos,
such as those hosting X-ray clusters.
This success inspires hope that a similarly concise 
description can be found for the hot gaseous component.
Given a population of dark matter halos from an N-body simulation
(or from some semi-analytic model such as
extended Press-Schechter), can we deduce
the global properties of the baryonic component inside each halo?
In this paper we have worked towards providing a 
prescription which is simple enough to apply
broadly while remaining physically well motivated.

In the model presented here,
the gas is assumed to initially have  energy per unit mass 
equivalent to that of the dark matter;  this energy can be
modified by removal of low entropy gas (to form
stars), addition of feedback energy expected from supernovae and 
accreting black holes, and mechanical work done as the gas
expands or contracts.  The gas is assumed to 
redistribute itself into a polytropic distribution in
hydrostatic equilibrium with the DM potential;
given the constraints on the total energy 
and the surface pressure at the virial radius, the gas
distribution is entirely specified.

We applied two variations of this method to a catalogue
of cluster-sized dark
matter halos drawn from a large cosmological N-body simulation.
In the first variant, the mass, virial radius, and
concentration of each halo
was measured, and the mass profile was assumed to follow 
a spherically symmetric NFW profile. 
To determine the gas distribution in this case means solving
Eqns. (\ref{eqn:eqen}) and  (\ref{eqn:eqpres}).
We then allowed for complex,
nonspherical profiles and substructure
by using the full set of particle positions and velocities
in each N-body halo to determine the potential and
kinetic energy.
These two methods give similar results, but assuming a
spherical NFW profile gives slightly lower temperatures
on average and gives significantly less scatter in the
mass-temperature and X-ray luminosity-temperature than is observed.

Simply assuming the gas follows the dark matter leads to too
low temperatures and too high central densities, since the DM profile
has a cusp.   The polytropic rearrangement increases the central
temperature while decreasing the density, removing the cusp.
Removing low entropy gas for star formation further increases
temperatures and reduces density.  However, neither of these processes
changes the self-similar nature of the model.
Including energy from feedback does change this, because
in massive clusters the energy input will be small compared to
the total binding energy, while for smaller masses it can
have more of an impact.  We also implemented a simple
approximation for including nonthermal pressure support;
including a relativistic component in this way leads to somewhat lower
temperatures and slightly higher densities.

Essentially two dimensionless numbers are required to prescribe
the state of the gas in a given DM potential:  the fraction
of gas mass transformed to a condensed (primarily stellar) form
(determined by observations to have the value $f_s\approx0.12$); 
and the feedback from the condensed component, for which
a plausible estimate for the energy output from 
supernovae and black holes that is trapped  in the cluster gas
is 
$\epsilon f_s \approx 0.12\cdot 3.9\times 10^{-5} = 4.7\times10^{-6}$.

\begin{figure}
\plotone{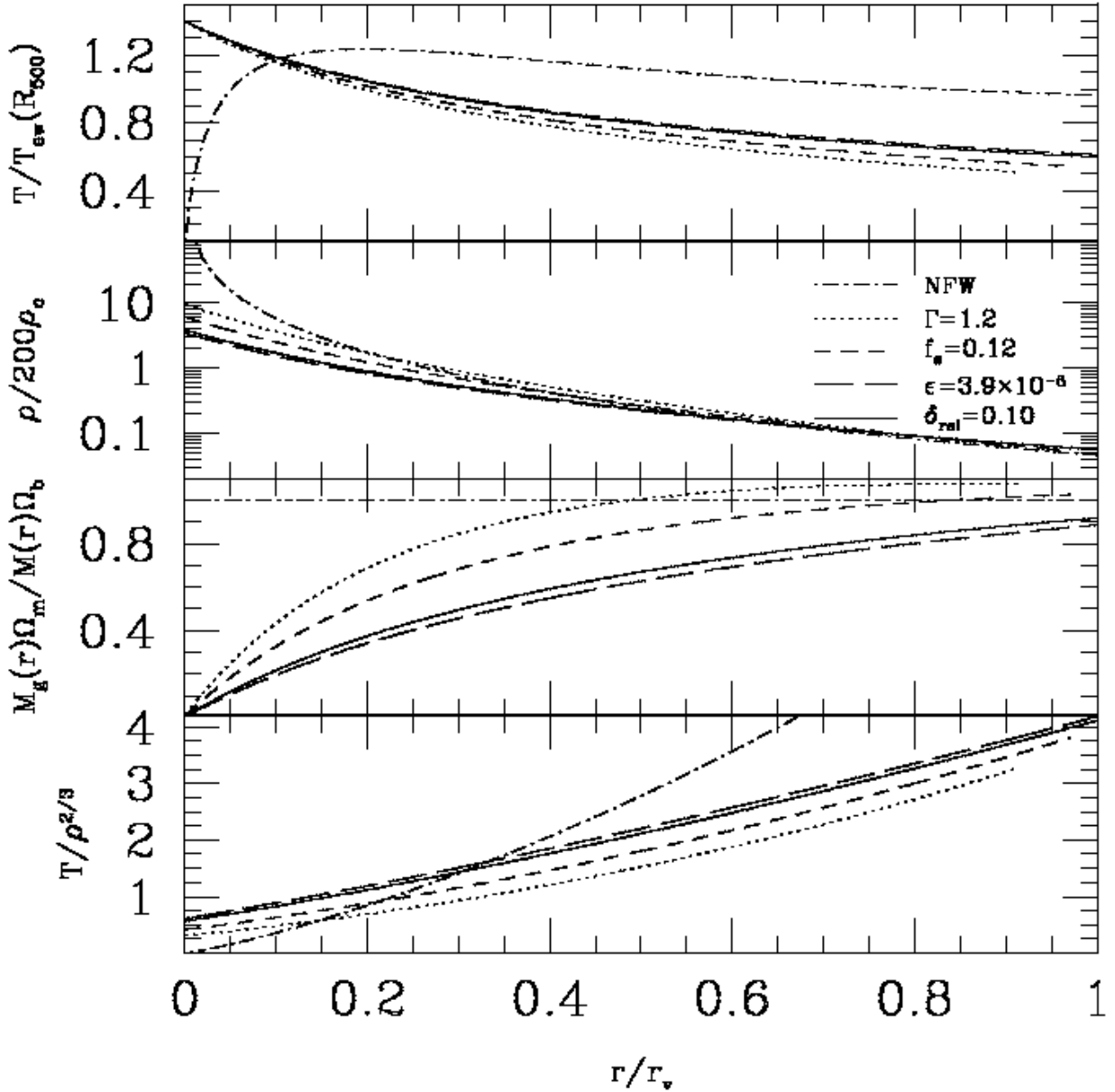}
\caption{Radial distributions of the temperature, density,
mass fraction, and entropy
of the gas for different model assumptions.
{\it Dot-dashed:} gas follows NFW profile.
{\it Dotted:} hydrostatic equilibrium with polytropic index $\Gamma=1.2$.
{\it Short-dashed:} also removing 12\% of the gas to form stars.
{\it Long-dashed:} also including energy input from feedback.
{\it Solid:} also with  a relativistic component.
In the bottom panel, the entropy is scaled by a factor
of $T_{ew}/(200\rho_c)^{2/3}$.
\label{fig:profs} }
\end{figure}

\begin{figure}
\plotone{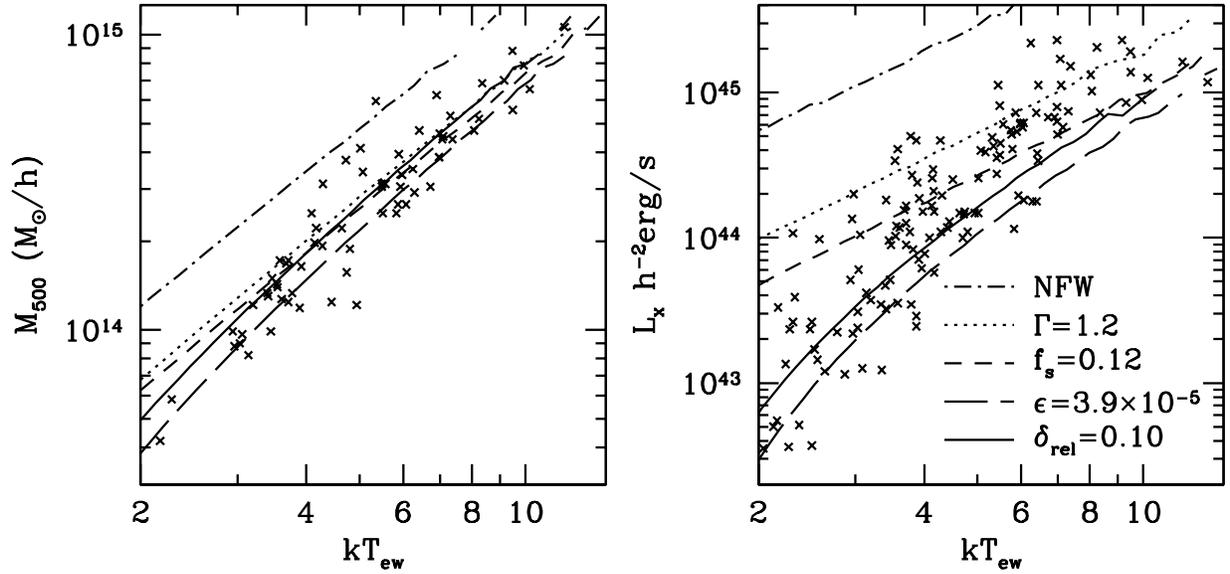}
\caption{ Results of the model applied to a population
of DM halos.  Line types as in Fig$.$\,\ref{fig:profs};
lines show the median value as a function of
emission-weighted $T$ inside overdensity 500.
{\it Left:} mass (measured at overdensity 500). 
{\it Right:} bolometric X-ray luminosity.
The data points in these two plots are from \citet{MBBPH04}.
\label{fig:rels} }
\end{figure}

\begin{figure}
\plotone{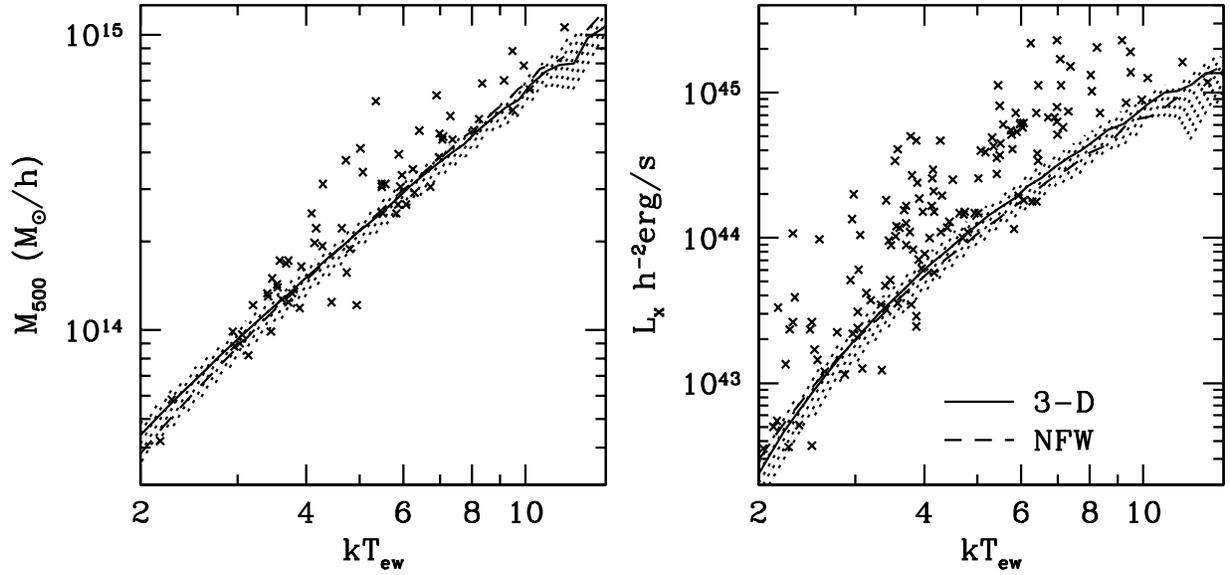}
\caption{ Results from using the directly computed
three-dimensional potential of
the simulated halos.  The
solid line in each panel is the median value, and the shaded region encloses
68\% of the clusters.  The corresponding NFW approximation is shown as
a dashed line.  Parameters used are $\Gamma=1.2, f_s=0.12,
\epsilon=3.9\times 10^{-5},$ and $\delta_{rel}=0$.
Points are the same as in Fig$.$\,\ref{fig:rels}.
\label{fig:3drels} }
\end{figure}

\begin{figure}
\plotone{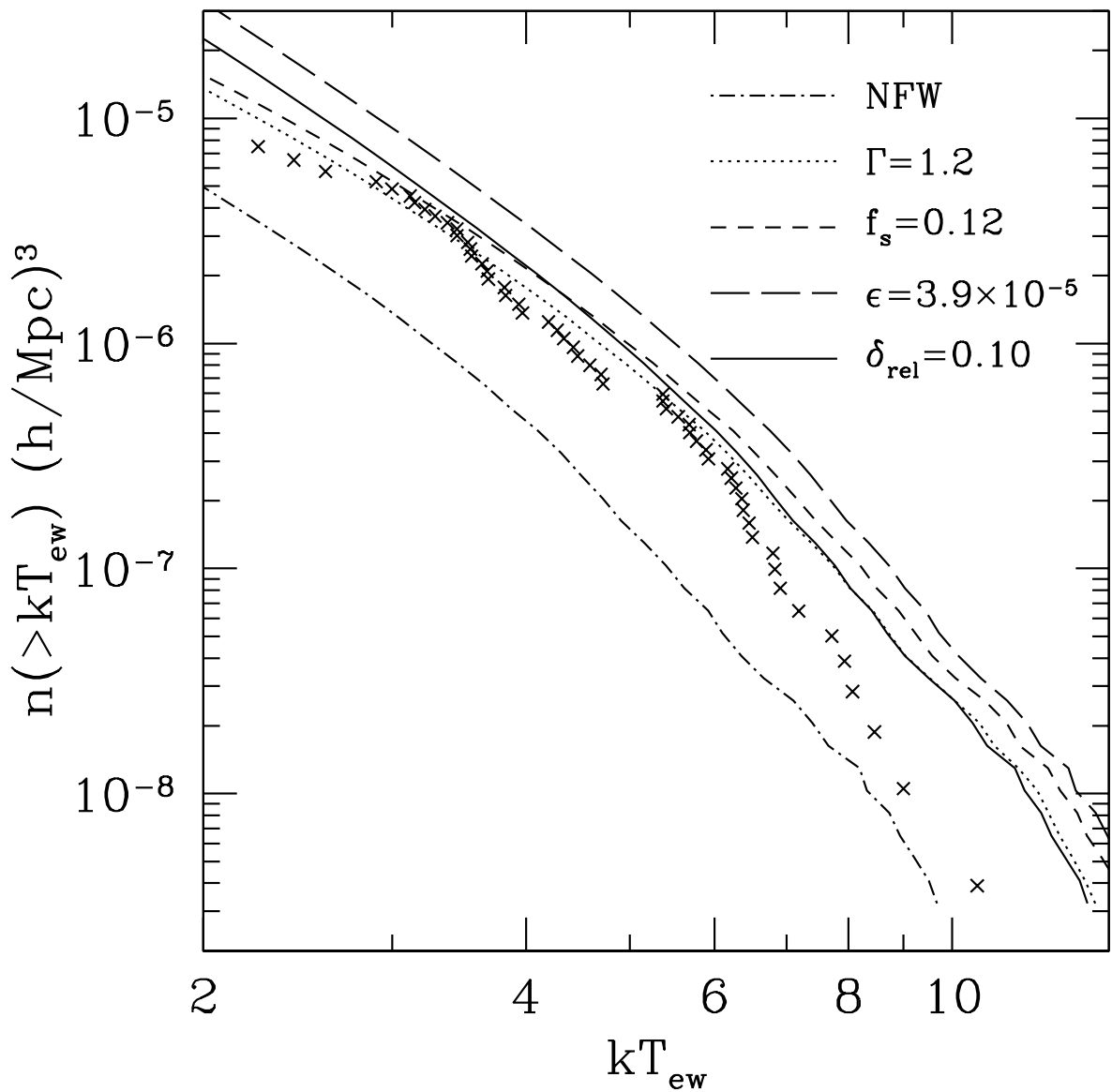}
\caption{ The cumulative temperature distribution function,
varying parameters as in Figs$.$\,\ref{fig:profs} and \ref{fig:rels}.  
(using the same line types);
data points are from \citet{IRBTK02}.
This function will be very sensitive to the choice of
cosmological model, particularly $\Omega_m$ and $\sigma_8$.
\label{fig:nofm} }
\end{figure}

The utility of fully understanding the properties
of the intergalactic medium in clusters can be seen
in Fig$.$\,\ref{fig:nofm}, which shows the
cumulative temperature function;  the
data points are from \citet{IRBTK02}.
The lines 
(the line types are the same as in 
Fig$.$\,\ref{fig:profs} and Fig$.$\,\ref{fig:rels})
demonstrate how, as 
different processes are included, the
resulting temperature function can change
quite dramatically.
(Note that the curves are the number density at $z=0$,
whereas at the highest $T$ objects are sufficiently 
rare such that the data points actually reflect the cluster
density at $z>0$, which is lower).
There is an apparent conflict in that no single model
seems to satisfy all the observations simultaneously
but this may not be a serious problem, and may
in fact reflect the strength of the models.  In order to 
reproduce the observed $M_{500}-T$ and $L_x-T$ relations
requires a high $T$ and low density, in other words a
significant amount of star formation and feedback.
But this seems to predict too high a number density 
of clusters for
a given T, as is seen in the plot of $n(>kT)$.  However,
rather than a problem with the gas model, this may simply
be due to our choice of cosmological parameters
$\Omega_m$ and $\sigma_8$ when generating the cluster
catalogue.  The mass function from our N-body simulation
is also too high when compared to that observed by SDSS 
\citep[see Fig$.$\,2 of][]{YBB05}.  Thus an accurate
mass-temperature relation would also lead to an overestimate
in the temperature function,
so the failure to reproduce the observed $n(>kT)$ relation
reflects a failure of the cosmological model; 
lowering $\sigma_8$ and/or 
$\Omega_m$ would alleviate this problem without
significantly altering the predicted $M_{500}-T$ and $L_x-T$ relations.
A 10\% reduction in $\sigma_8$ would reduce the number
of clusters with $kT>4$ by roughly a factor of two.
This points out the usefulness of the cluster number
density as a probe of cosmological parameters,
but also the necessity of including all the relevant
physics accurately.

A variety of telescopic surveys in many wavelength bands
will soon greatly multiply the number of galaxy clusters catalogued,
particularly at higher redshifts.  Unlocking the power of
these new observational datasets as cosmological probes
will require sophisticated theoretical predictions. 
In the future we plan to apply the methods developed here
to explore the properties of clusters at higher redshifts
and make detailed predictions for X-ray and SZE surveys
in many different cosmological models.

\begin{figure}
\plotone{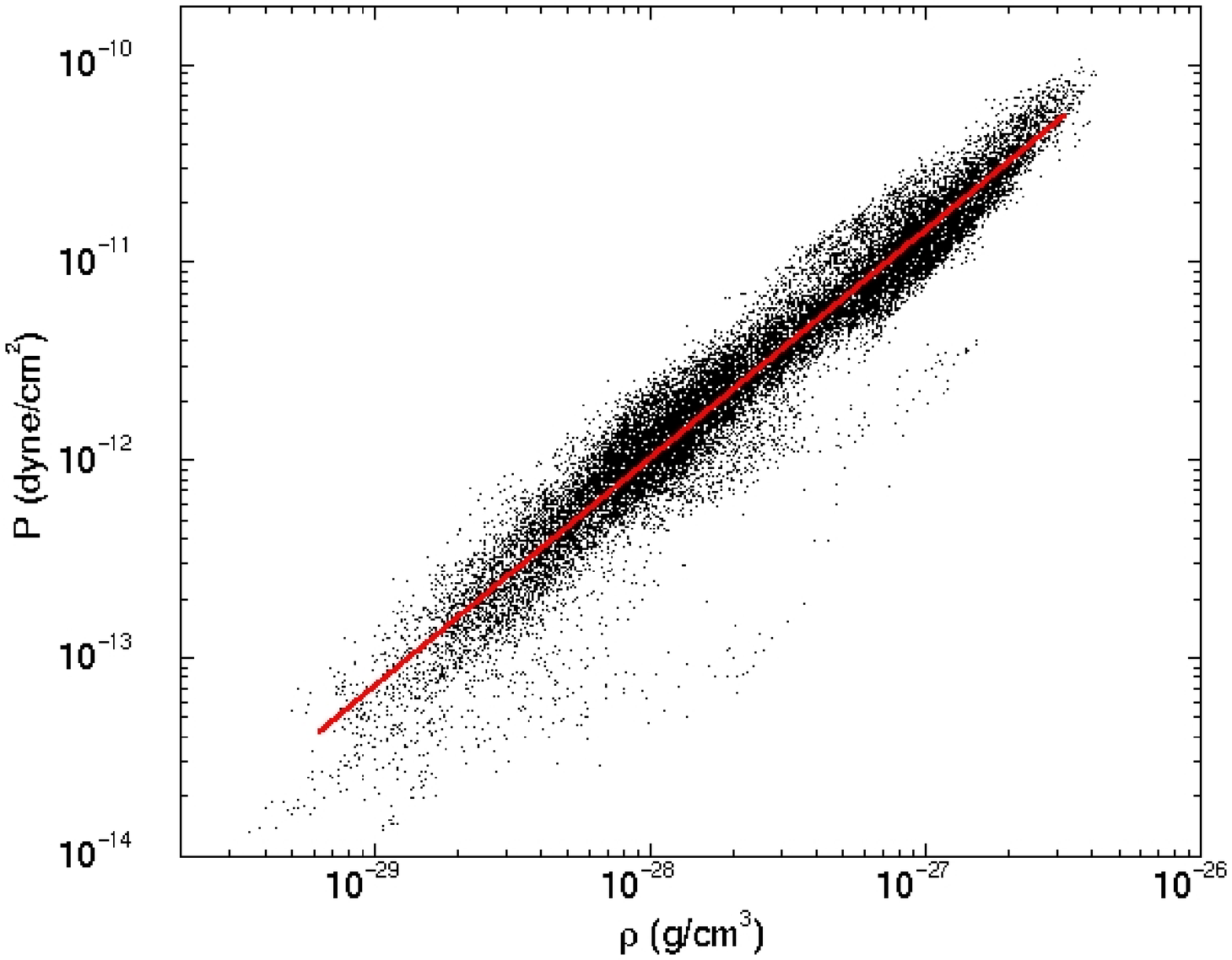}
\caption{Pressure as a function of density, from an
adiabatic simulation of a massive cluster by Greg Bryan.
The line has a logarithmic slope of $\Gamma$=1.15. 
\label{fig:gbrvrho} }
\end{figure}

\acknowledgments
Many thanks to Greg Bryan for kindly supplying
Fig$.$\,\ref{fig:gbrvrho}, and to Ian McCarthy for helpful discussions
and for making available the observational data 
from \citet{MBBPH04}, as well as the anonymous referee for
a careful and helpful reading of the manuscript.
Computations were performed on the National Science Foundation
Terascale Computing System at the Pittsburgh Supercomputing Center,
with support from NCSA under
NSF Cooperative Agreement ASC97-40300, PACI Subaward 766.
Computational facilities at Princeton were provided by
NSF grant AST-0216105.
Research support for AB comes from the Natural Sciences and Engineering
Research Council (Canada)
through the Discovery grant program.   AB would also like to acknowledge
support from the Leverhulme
Trust (UK) in the form of the Leverhulme Visiting Professorship at the
University of Oxford.




\end{document}